\input{aipcheck}

\documentclass[
    ,final            
  ]
  {aipproc}

\layoutstyle{6x9}

\newcommand{\kb}{\underline{k}}
\newcommand{\rb}{\underline{r}}
\newcommand{\xb}{\bar{x}}
\newcommand{\vb}{\bar{v}}
\begin{document}
\hspace{11cm} LPT-08-90

\hspace{11cm} CPHT-PC086.1108
\vspace{-1cm}

\title{ $\gamma^{*}\to\rho_{T}$ impact factor with twist three accuracy}

\author{I.V. Anikin}{
  address={Bogoliubov Laboratory of Theoretical Physics, JINR,
             141980 Dubna, Russia}}

\author{D.Yu. Ivanov}{
  address={Sobolev Institute of Mathematics and Novosibirsk State University,
630090 Novosibirsk, Russia}}

\author{B. Pire}{
  address={Centre  de Physique Th{\'e}orique, \'Ecole Polytechnique, CNRS,
   91128 Palaiseau, France}}

\author{L. Szymanowski}{
  address={Soltan Institute for Nuclear Studies, Warsaw, Poland}}

\author{S. Wallon}{
  address={ LPT, Universit\'e d'Orsay, CNRS, 91404 Orsay, France}}

\begin{abstract}
We evaluate the impact factor of the transition $\gamma^{*} \to \rho_{T}$ taking into
 account the twist $3$ contributions. We show that
a gauge invariant expression is obtained with the help of QCD equations of motion. Our results are free of end-point singularities. This opens the way to a
consistent treatment of factorization for exclusive processes with a transversally polarized vector meson.
\end{abstract}

\maketitle


\section{Motivation}

The study of exclusive reactions in the generalized Bjorken regime has been the scene of significant
progresses in the recent years, thanks to the factorization properties of the leading twist amplitudes \cite{fact}
for deeply virtual Compton scattering and deep exclusive meson production. It however turned out that
transversally polarized rho meson production did not enter the leading twist controlable case \cite{DGP} but only
the twist 3 more intricate part of the amplitude \cite{MP,AT}. An understanding of the quark-gluon structure of a
transversally polarized vector meson is however an important task of hadronic physics if one cares about studying
confinement dynamics. This quark gluon structure may be described by distribution amplitudes which have been
discussed in great detail \cite{BB}. On the experimental side, a continuous effort has been devoted to the exploration of
 $\pi \pi$ photo and electro-production, from moderate to very large energy. The kinematical analysis of the final
 meson pair allows then to separate the different helicity amplitudes, hence to measure the transversally polarized $\rho$
 meson production amplitude. Although non-dominant for deep electroproduction, this amplitude is by no means negligible
 and needs to be understood in terms of QCD. Up to now, experimental information comes from electroproduction on a proton
 or nucleus. Future progress may come from real or virtual photon photon collisions, which may be accessible either at electron
 positron colliders or in ultraperipheral collisions at hadronic colliders, as recently discussed \cite{IP,PSW}.

\section{Calculation of the impact factor}
We are interested in  the description of high energy reactions such as
\begin{eqnarray}
\label{prgg}
\gamma^*(q)+\gamma^*(q^\prime)\to \rho_T(p_1)+\rho(p_2)
\end{eqnarray}
or
\begin{eqnarray}
\label{prgP}
\gamma^*(q)+N\to \rho_T(p_1)+N
\end{eqnarray}
where the virtual photons carry large squared  momenta $q^2=-Q^2$ 
($q'^2=-Q'^2$) $\gg \Lambda^2_{QCD}$\,, and
the Mandelstam variable $s$ obeys the condition
$s\gg Q^2,\,Q^{\prime\, 2}, -t \simeq \rb^2$. 
Neglecting meson masses, one considers for reaction (\ref{prgg}) the light cone vectors  $p_1$ and $p_2$
as the vector meson momenta  ($2\,p_1\cdot p_2=s$);
the "plus" light cone direction is directed along $p_1$ and
 the ``minus'' light cone direction along $p_2$ with vector $n$ 
 defined as $p_2/(p_1 \cdot p_2)$. In this Sudakov light-cone basis, transverse euclidian momenta
are denoted with underlined letters. The virtual photon momentum $q$  reads
$ q=p_1-\frac{Q^2}{2}\, n$.
The impact representation of the scattering amplitude for the reaction (\ref{prgg})  is
\begin{eqnarray}
\label{BFKLamforward}
{\cal M}=\frac{i s}{(2\pi)^2}\!\!
\int\frac{d^2\kb}{\kb^2} \Phi^{ab}_1(\kb,\,\rb-\kb) \!\!
\int\frac{d^2\kb'}{\kb'^2} \Phi^{ab}_2(-\kb',\,-\rb+\kb') \!\!\!\!\!
\int\limits_{\delta-i\infty}^{\delta+i\infty} \frac{d\omega}{2\pi i}
\biggl(\frac{s}{s_0}\biggr)^\omega G_\omega (\kb,\kb',\rb) 
\end{eqnarray}
where  $G_\omega$ is the 4-gluons Green function which obeys the BFKL equation. $G_\omega$ reduces to $1/\omega \,\delta(\kb -\kb') \kb^2/(\rb-\kb)^2$ within 
Born approximation. 
We focus here on the $\gamma^* \to \rho $ impact factor $\Phi$  of the  subprocess
\begin{eqnarray}
\label{ggsubpr}
   g(k_1,\varepsilon_{1})+\gamma^*(q)\to g(k_2, \varepsilon_{2})+\rho_T(p_1) \,,
\end{eqnarray}
which is the integral of the $\kappa$-channel discontinuity of the  S-matrix element  
 ${\cal S}^{\gamma^*_T\, g\to\rho_T\, g}_\mu\,:$
\begin{eqnarray}
\label{imfac}
\Phi^{\gamma^*\to\rho}(\kb,\,\rb-\kb)= e^{\gamma^*\mu}\, \frac{1}{2s}\int\frac{d\kappa}{2\pi}
\, \hbox{Disc}_\kappa \,  {\cal S}^{\gamma^*\, g\to\rho\, g}_\mu(\kb,\,\rb-\kb)\,,
\end{eqnarray}
where $\kappa=(q+k_1)^2$ denotes the Mandelstam variable $s$ for the subprocess (\ref{ggsubpr}). Note that the two reggeized 
gluons have so-called non-sense polarizations $\varepsilon_1=\varepsilon_2^*=p_2\sqrt{2/s}\,.$ Considering now the forward limit for simplicity, 
the gluon momenta read
\begin{eqnarray}
\label{gmom}
k_1=\frac{\kappa+Q^2+\kb^2}{s} p_2 + k_T, \quad
k_2=\frac{\kappa+\kb^2}{s} p_2 + k_T, \quad
k_1^2=k_2^2=k^2_T=-\kb^2\,.
\end{eqnarray}

\begin{figure}[htb]
\begin{tabular}{cccc}
\hspace{.5cm}\includegraphics[width=0.275\textwidth]{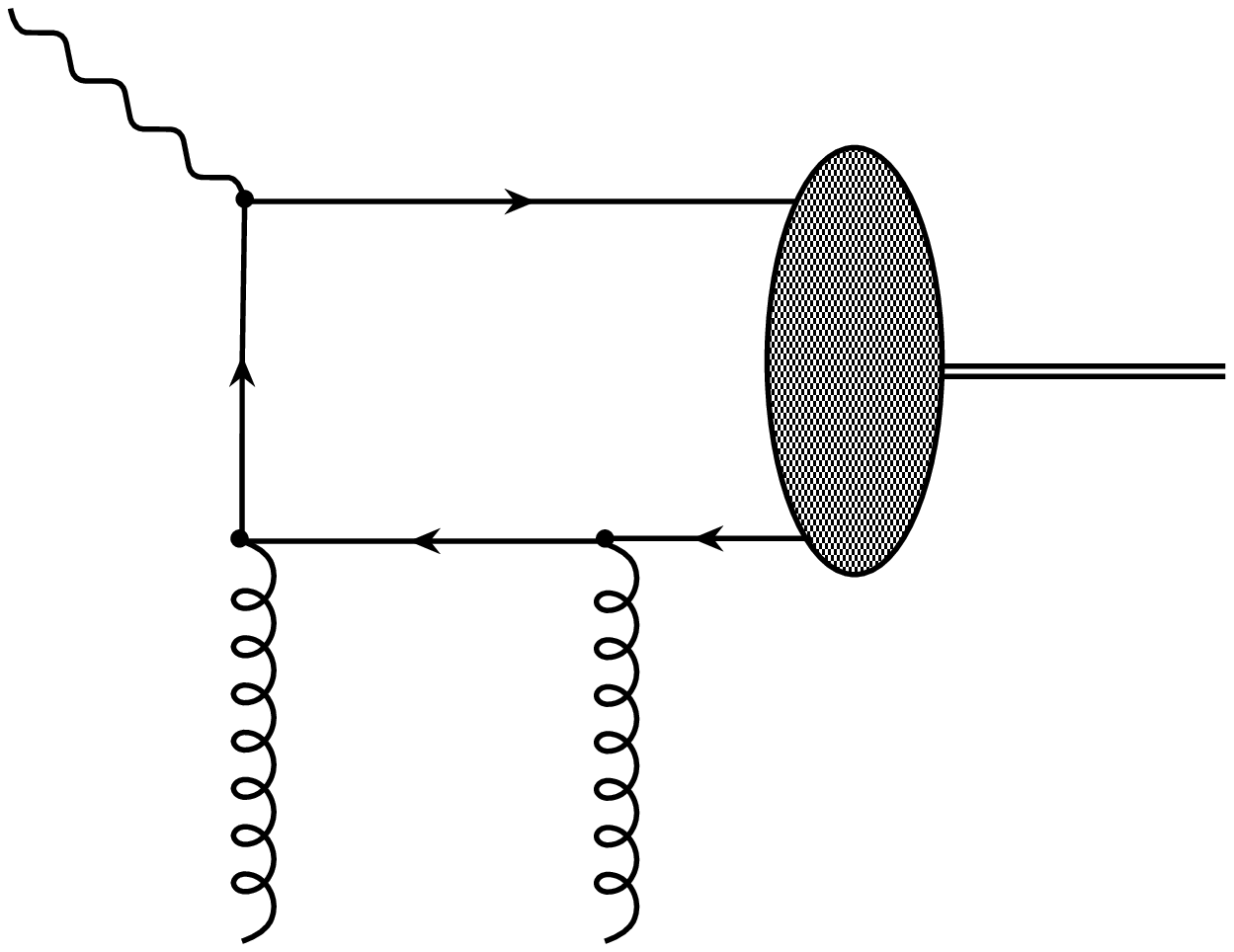}
&\hspace{-1.3cm}
\includegraphics[width=0.275\textwidth]{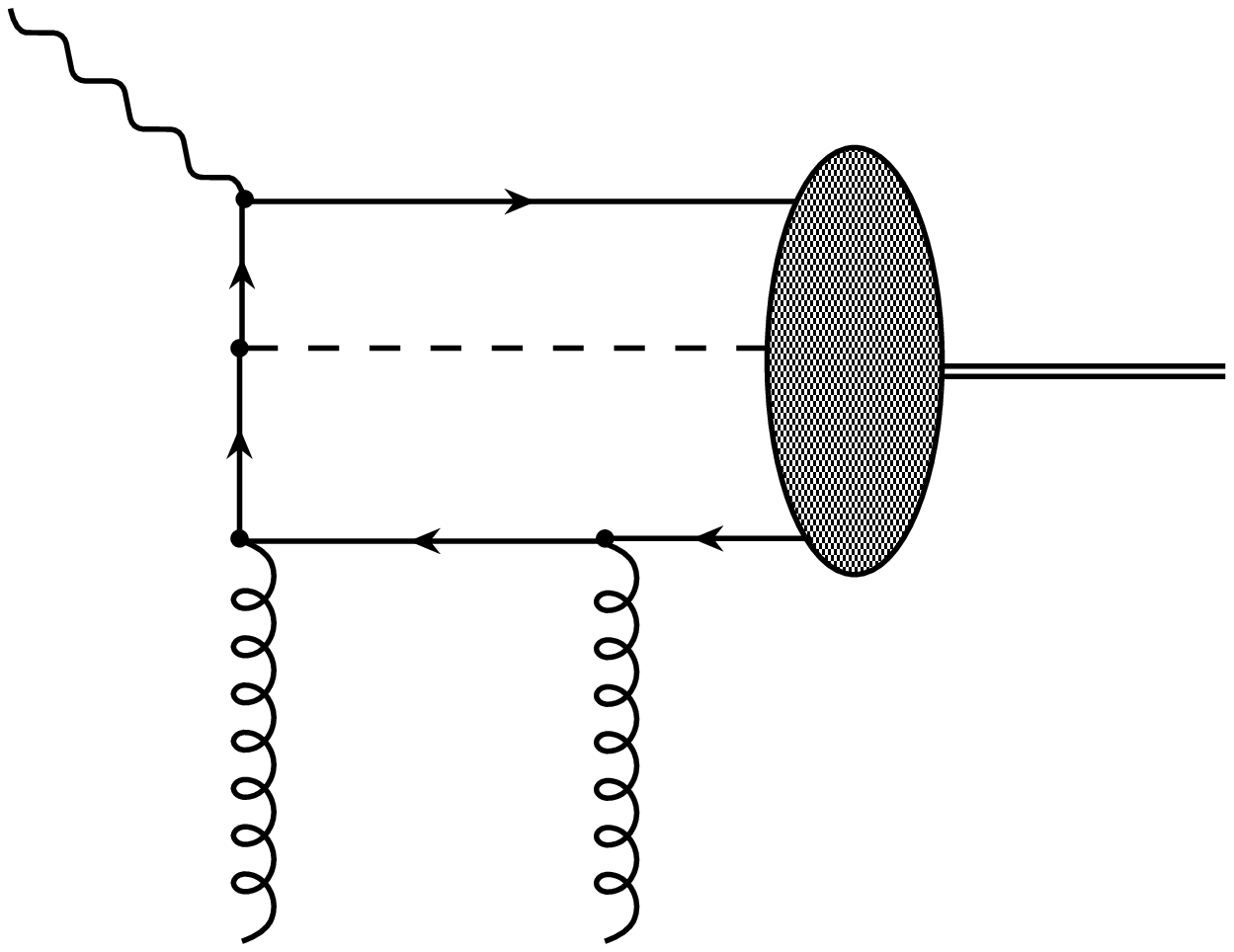} &
\hspace{-1.3cm}
\raisebox{-.1cm}{\includegraphics[width=0.275\textwidth]{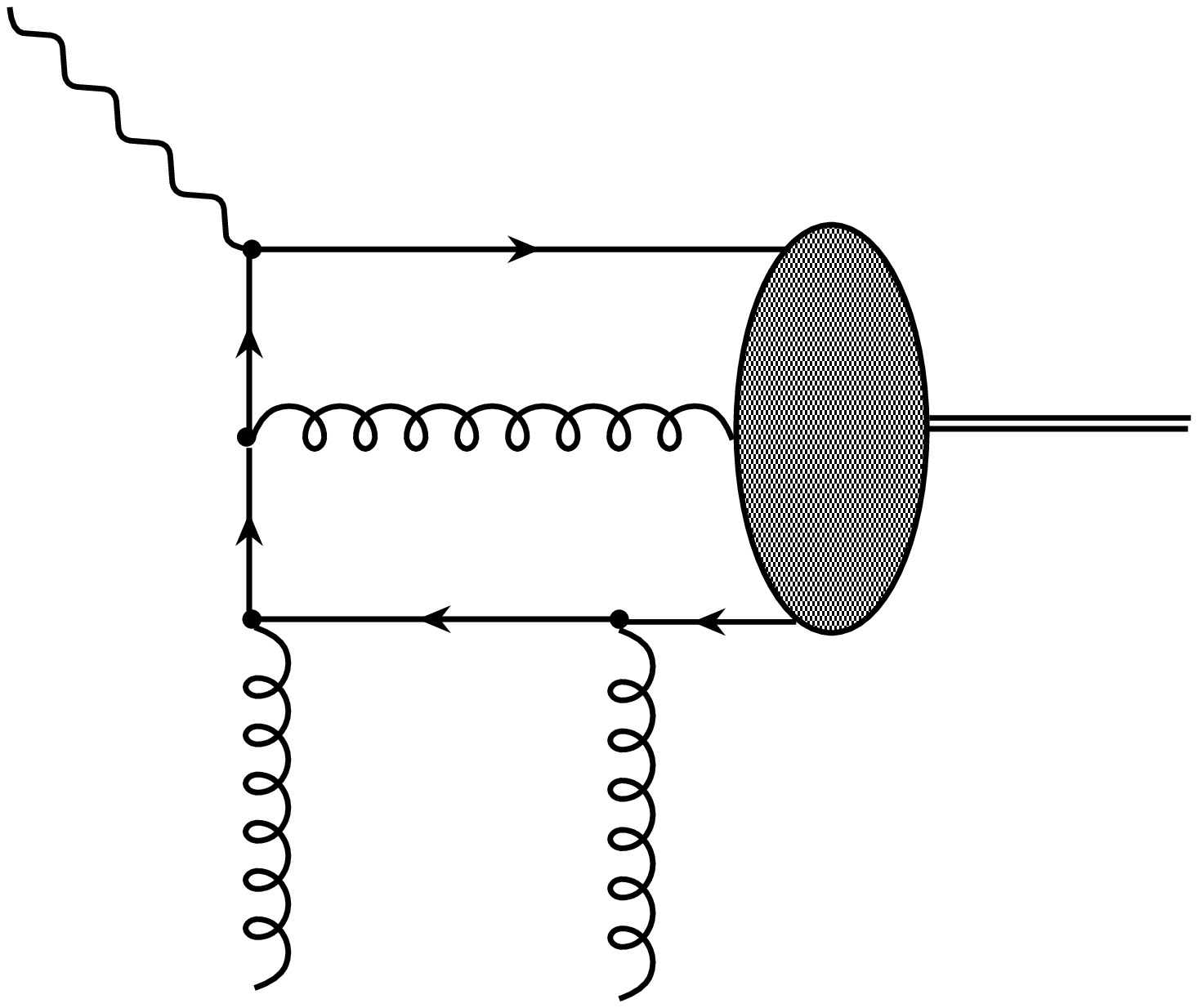}}
&
\hspace{-1.3cm}
\includegraphics[width=0.275\textwidth]{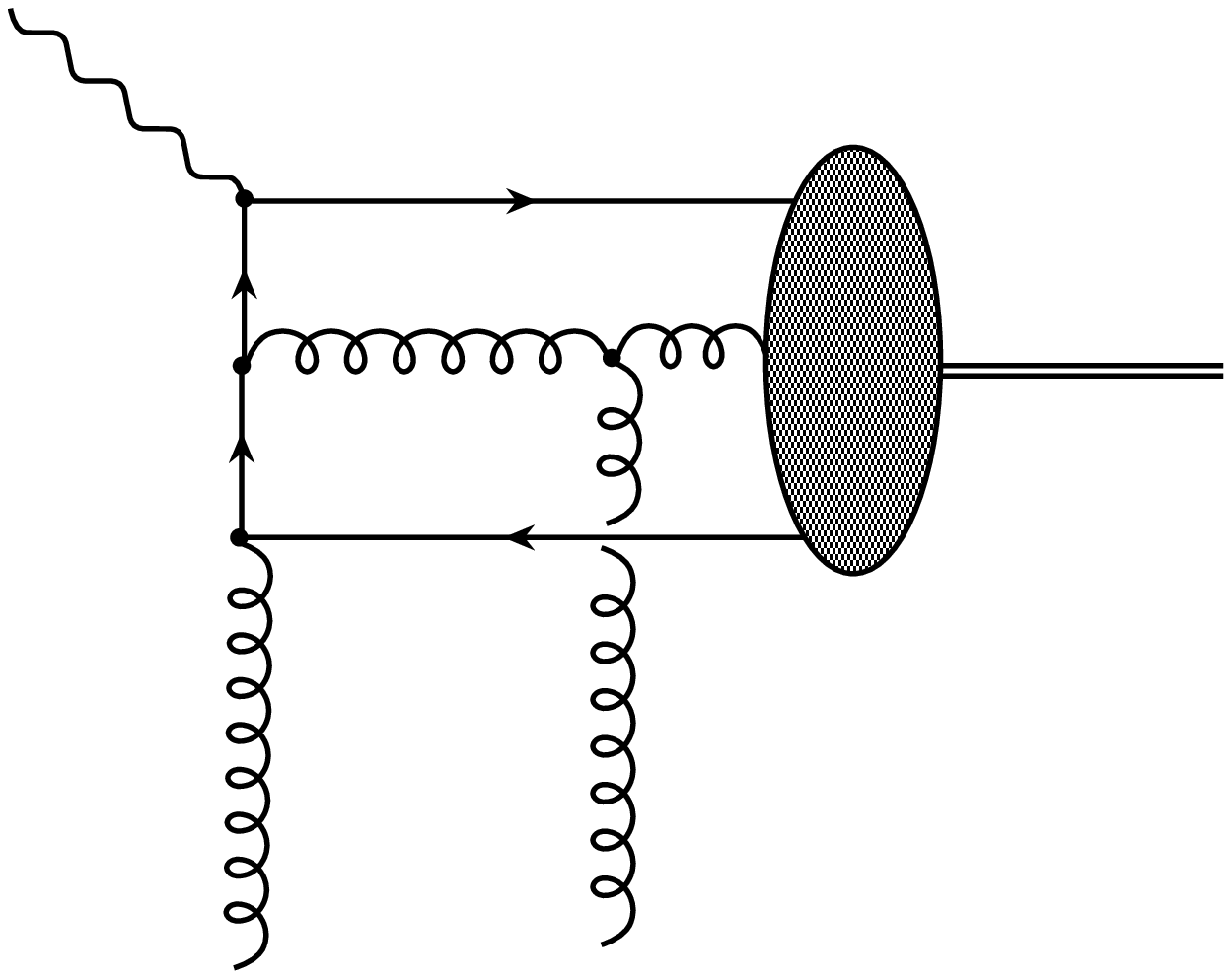}\vspace{-.25cm}
\\
\vspace{-.1cm}
a & \hspace{-1.1cm} b & \hspace{-1.3cm}c & \hspace{-1.5cm}d
\end{tabular}
 \vspace{-1cm}
\caption{Contributions to the impact factor with the quark-antiquark correlators without transverse derivatives (a), 
 with transverse derivatives  (b), and with the quark-antiquark-gluon correlators (c, d).
   \label{Fig1}}
\end{figure}

The impact factor $\Phi$  can be calculated within the
collinear factorization. It is a convolution of
 perturbatively calculable hard-scattering amplitudes $H$ and soft correlators $S$ involving relevant $\rho$-meson
distribution amplitudes, symbolically written as
\begin{equation}
\label{factorisation}
\Phi= \int d^4l \cdots \, \hbox{tr} [H(l \cdots) \, S(l \cdots)]\,,
\end{equation}
integrating over the $n$ ($n \ge 2$)  body phase-space.
 Working within the  light-cone collinear factorization framework, let us first derive the contribution of the diagrams with the quark-antiquark
correlators. The basic tool is to decompose any hard  coefficient function $H(l)$  around
a dominant ``plus" direction:
\begin{eqnarray}
\label{expand}
H(l) = H(x \,p) + \frac{\partial H(l)}{\partial l_\alpha} \biggl|_{l=x \,p}\biggr. \, (l-x \,p)_\alpha + \ldots \,,
\quad        \,\,\,\, (l-x \,p)_\alpha \approx l^T_\alpha ,
\end{eqnarray}
where the first term is associated with the usual twist-$2$ contribution while the term with the derivative with
the kinematical twist $3$. After Fourier transform, this second hard term
is accompanied in (\ref{factorisation}) by a soft correlator involving  derivatives of the fields.
These contributions are shown on  Fig.\ref{Fig1}a (no  transverse derivatives),
and  Fig.\ref{Fig1}b (with transverse derivatives).\vspace{-.45cm}

This  allows us to factorize the amplitude in the momentum
space. After a Fierz decomposition of the Dirac matrices
$\psi (0) \, \bar\psi(z)$ and $\psi (0) \, i \, \stackrel{\longleftrightarrow}
{\partial_{\perp}}\, \bar\psi(z)$, the amplitude takes, up to higher twist terms, the simple factorized form symbolically
written as:
\begin{equation}
\Phi \sim
\int dx \,\left\{\hbox{tr} \left[ H_{q \bar{q}}(x\, p ) \, \Gamma \right] \, S_{q \bar{q}}^\Gamma (x) +\hbox{tr} 
\left[ \partial_{\perp} H_{q \bar{q}}(x\, p  ) \, \Gamma \right] \,  \partial_{\perp}S_{q \bar{q}}^\Gamma (x )
\right\}  
\end{equation}
where $\Gamma$ stands for some $\gamma$-matrix from a full set
and the soft part  reads symbolically
\begin{eqnarray}
\label{soft}
S_{q \bar{q}}^\Gamma (x) &=& \int \frac{d\lambda}{2\pi} \, e^{-i\lambda x}
\langle \rho(p) | \bar\psi(\lambda n)\,\Gamma \,\psi(0)| 0 \rangle \nonumber \\
 \partial_{\perp}S_{q \bar{q}}^\Gamma (x)&=& \int \frac{d\lambda}{2\pi} \, e^{-i\lambda x}
\langle \rho(p) | \bar\psi(\lambda n)\,\Gamma \, i \, \stackrel{\longleftrightarrow}
{\partial_{\perp}}\, \psi(0)| 0 \rangle \,.
\end{eqnarray}
Within twist 3 order, the  correlators which need to be parameterized are thus\footnote{In the following definitions, ${\cal F}_1$ and ${\cal F}_2$ denote 2 and 3-body relative Fourier transforms along
$p.$} \cite{AT}
\begin{eqnarray}
\label{par1v}
&&\langle \rho(p)|\bar\psi(z)\gamma_{\mu} \psi(0)|0\rangle
\stackrel{{\cal F}_1}{=} m_\rho \, f_\rho \,[
\varphi_1(x)\, (e^*\cdot n)p_{\mu}+\varphi_3(x)\, e^{*T}_{\mu}]\,,
\\
\label{par1a}
&&\langle \rho(p)|
\bar\psi(z)\gamma_5\gamma_{\mu} \psi(0) |0\rangle
\stackrel{{\cal F}_1}{=}m_\rho \, f_\rho \,
i\varphi_A(x)\,
\varepsilon_{\mu\lambda\beta\delta}\,
e^{*T}_{\lambda}\, p_{\beta} \, n_{\delta}
\end{eqnarray}
and, for the quark-antiquark operators with the transverse derivatives,
\begin{eqnarray}
\label{par1.1v}
&&\langle \rho(p)|
\bar\psi(z)\gamma_{\mu}
i\stackrel{\longleftrightarrow}
{\partial^T_{\alpha}} \psi(0)|0 \rangle
\stackrel{{\cal F}_1}{=}m_\rho \, f_\rho \,
\varphi_1^T(x) \, p_{\mu} e^{*T}_{\alpha} \, ,
\\
\label{par1.1a}
&&\langle \rho(p)| \bar\psi(z)\gamma_5\gamma_{\mu}
i\stackrel{\longleftrightarrow}
{\partial^T_{\alpha}} \psi(0) |0\rangle
\stackrel{{\cal F}_1}{=}m_\rho \, f_\rho \,
i \, \varphi_A^T (x) \, p_{\mu} \,
\varepsilon_{\alpha\lambda\beta\delta}
\, e^{*T}_{\lambda} \, p_{\beta} \, n_{\delta}\,.
\end{eqnarray}
A careful treatment of the correlators needs to take into account the equations of 
motion for the fermion field which read $\langle i  \stackrel{\rightarrow}
{\hat D}(0) \psi(0)\, \bar \psi(z)\rangle = 0$, and the corresponding one for the antifermion field, 
where $i\stackrel{\rightarrow}{D}_\mu=i\stackrel{\rightarrow}{\partial}_\mu
+g \, A_\mu$.  This shows that one must also consider 
 diagrams such as the one shown on Fig. \ref{Fig1}c, d, with  the three body quark-antiquark-gluon nonlocal operators. One
parameterizes these new non-perturbative objects as
\begin{eqnarray}
\label{par1.2}
&&\langle \rho(p)|
\bar\psi(z_1)\gamma_{\mu}g A_{\alpha}^T(z_2) \psi(0) |0\rangle
\stackrel{{\cal F}_2}{=}m_\rho \, f_3 \,
B(x_1,x_2)\, p_{\mu} e^{*T}_{\alpha},
\nonumber\\
&&\langle \rho(p)|
\bar\psi(z_1)\gamma_5\gamma_{\mu} g A_{\alpha}^T(z_2) \psi(0) |0\rangle
\stackrel{{\cal F}_2}{=}m_\rho \, f_3 \,
i \,
D(x_1,x_2)\, p_{\mu}
\varepsilon_{\alpha\lambda\beta\delta}
\, e^{*T}_{\lambda} \, p_{\beta} \, n_{\delta}\,,
\end{eqnarray}
where fractions $x_1$, $\bar x_2$ and $x_2-x_1$ correspond to the quark, antiquark and gluon.
We 
decompose the impact factor as the sum of non-flip  ($\Phi_{nf}$) and flip  ($\Phi_{f}$) contributions
\begin{equation}
\label{impactNonFlipFlip}
\Phi^{\gamma^*_T\to\rho_T}(\kb^2)= \Phi_{nf}^{\gamma^*_T\to\rho_T}(\kb^2) \,(-e_{\gamma^*} \cdot e^*)+  \Phi_{f}^{\gamma^*_T\to\rho_T}(\kb^2) \!\!\left[ \frac{(e_{\gamma^*} \cdot k)(e^* \cdot k)}{\kb^2}+\frac{(e_{\gamma^*} \cdot e^*)}{2}\!\right].
\end{equation}
The gauge invariance of the total amplitude is guaranteed if $ \Phi(\kb^2=0) = 0$.
This is satisfied when using  
the equations of 
motion.
The procedure  is quite intricate and will be discussed in detail in a forthcoming publication.
We stress that the 
impact factor is gauge-invariant 
{\em provided} the  three body correlator contribution is taken into account. 
We only display
the results obtained in the Wandzura-Wilczek approximation 
\begin{eqnarray}
\hspace{0cm}\Phi_{nf\,(WW)}^{\gamma^*_T\to\rho_T}(\kb^2) \!\!\!\!\!\!
 &=&\!\!\!\!\!\!-\frac{e \, m_\rho f_\rho}{\sqrt{2}\, Q^2}\frac{\delta^{ab}}{2 N_c} \int\limits^1_0 dx\,\,
 \frac{\kb^2
  \left(\kb^2+2 \, Q^2 \, x \xb \right)}{
   \left(\kb^2+Q^2 \, x\, \xb \right)^2} \!
\left[\int\limits^x_0 dv 
 \frac{\varphi_1\left(v\right)}{\vb}+\int\limits^1_x dv 
 \frac{\varphi_1\left(v\right)}{v}\right]\!\!,\,\,\,\,\,\,\,\,\,\,\,
\\
\Phi_{f\,(WW)}^{\gamma^*_T\to\rho_T}(\kb^2)\!\!\!\!\!\!
&=&\!\!\!\!\!\!\frac{e \, m_\rho f_\rho}{\sqrt{2}\, Q^2} \frac{\delta^{ab}}{2 N_c}\int\limits^1_0 dx \, \frac{2 \,\kb^2\,Q^2}{\left( \kb^2 + Q^2 \,{x}\,\xb \right) ^2}\!
\left[\xb^2\! \int\limits^x_0 dv 
 \frac{\varphi_1\left(v\right)}{\vb}+x^2 \!\int\limits^1_x dv 
 \frac{\varphi_1\left(v\right)}{v}\right]\!\!,\,\,\,\,\,\,\,
\end{eqnarray}
where $\xb=1-x$.
Our results are free of end point singularities, thanks to the presence of the transverse momentum $\kb$ in the
denominators which regulates them.
\begin{theacknowledgments}
\noindent
 This work is partly supported by the ECO-NET program, contract 
18853PJ, the French-Polish scientific agreement Polonium, the grant 
ANR-06-JCJC-0084, the RFBR (grants 06-02-16215,
 08-02-00334, 08-02-00896) and 
the Polish Grant N202 249235.
\end{theacknowledgments}

\end{document}